\documentclass[useAMS,usenatbib,usegraphicx]{mn2e}

\usepackage{verbatim}
\usepackage{multirow}

\newcommand{\aj}{AJ}
\newcommand{\apj}{ApJ}
\newcommand{\apjl}{ApJL}
\newcommand{\mnras}{MNRAS}
\newcommand{\aap}{A\&A}

\newcommand{\pasj}{PASJ}
\newcommand{\araa}{ARAA}

\newcommand{\aapr}{A\&AR}

\newcommand{\lhls}{$\ell_h/\ell_s$}
\newcommand{\rxte}{{\it RXTE}}
\newcommand{\gro}{{GRO J1655-40}}
\newcommand{\gx}{{GX 339-4}}
\newcommand{\na}{New Astronomy}

\title[GBH hard state X-ray radiation mechanism change]
{Evidence for a change in the X-ray radiation mechanism in the
hard state of Galactic black holes}
\author[M. A. Sobolewska et al.]{M. A.
Sobolewska$^{1,2,3}$\thanks{E-mail:
  msobolewska@cfa.harvard.edu}, I. E. Papadakis$^{2,3}$, C. Done$^{4}$,
J. Malzac$^{5,6}$\\
$^1$ Harvard-Smithsonian Center for Astrophysics, 60 Garden Street,
  Cambridge, MA 02138, USA\\
$^2$Foundation for Research and Technology - Hellas, IESL, Voutes, 71110
  Heraklion, Crete, Greece\\ 
$^3$University of Crete, Department of Physics and Institute
of Theoretical \& Computational Physics, Voutes, 71003 Heraklion,
  Crete, Greece\\ 
$^4$University of Durham, Department of Physics, South Road, DH1 3LE,
  Durham, UK\\
$^5$ Universit\'e de Toulouse; UPS-OMP; IRAP;  Toulouse, France\\
$^6$ CNRS; IRAP; 9 Av. colonel Roche, BP 44346, F-31028 Toulouse cedex 4, France}
\begin{document}

\bibliographystyle{plainnat}
\date{}

\pagerange{\pageref{firstpage}--\pageref{lastpage}} \pubyear{2010}

\maketitle

\label{firstpage}

\begin{abstract}
We present results on spectral variability of two Galactic black hole
X-ray binaries, \gro\ and \gx, in the hard state. We confirm a
transition in behaviour of the photon index with
luminosity, such that the well known decrease in X-ray photon index
with decreasing luminosity only continues down to $L_{\rm bol} \sim
0.01L_{\rm E}$. Below this point the photon index increases again.
For Comptonisation models, this implies that the ratio of the Compton
luminosity to seed photon luminosity, \lhls, changes with bolometric
luminosity, consistent with a scenario where seed photons change from
cyclo-synchrotron at the lowest luminosities to those from a truncated
disc. Alternatively, the transition could mark the point below which
the non-thermal jet starts to dominate, or where reprocessed photons
replace the viscous ones in an outflowing corona model.
\end{abstract}

\section{Introduction}

Galactic black hole binaries (GBHs) are strong sources of X-ray radiation. This
radiation can be decomposed into two main components: soft thermal
radiation originating in an accretion disc, and hard power-law like
emission, probably originating in the process of Comptonisation of the
disc photons by hot electrons located in the so-called corona (see
reviews by Remillard \& McClintock 2006; Done, Gierli\'nski \&
Kubota 2007; and references therein).
The disk and corona contribute at different levels to the
total spectrum. Consequently, a number of different GBH spectral
shapes have been observed and classified into two main spectral states. In the soft
state the hot disc dominates the spectrum up to a few keV and is accompanied 
by a weak power-law tail with a soft photon index, $\Gamma>2$. In the hard
state the disc is much cooler and the total X-ray band is dominated by the 
radiation of the corona with typically a hard photon index, $\Gamma
< 2$.

The physical origin of the transitions between the spectral states is not clear
yet. Nevertheless, phenomenologically the evolution of the X-ray spectral shape
within a given state and during the transition from one state to another is
relatively well understood. The GBHs in outburst
track the same, continuous q-shaped pattern in the so-called
hardness-intensity diagram (HID, e.g. Fender, Belloni \& Gallo 2004; Dunn et al.
2008). From this diagram it follows that initially the X-ray
spectra of GBHs are
dominated by the hard colour and their intensity is low; the sources
are in a hard state and reside in the bottom/right corner of the HID.
With
time, the intensity increases but the spectra remain dominated by the
hard colour; the sources move upwards in the diagram. At some point,
further increase in intensity results in the decrease of hardness, and
the sources move toward the top/left part of the HID. Eventually, a
transition to a soft state takes place, resulting in populating the
top/left section of the HID. Towards the decay of the outburst the
intensity decreases and a transition back to a hard state is observed.
The soft-to-hard state transition occurs typically at lower intensity
than the hard-to-soft state transition at the beginning of the
outburst (e.g. Gierli\'nski \& Newton 2006), which leads to
a hysteresis pattern in the HID. Theoretical explanation of the
hysteresis
include effects of different source of coronal irradiation in the hard
and soft states (Meyer-Hofmeister, Liu \& Meyer 2005, 2009), and in
the most extreme cases a non-steady state behaviour when the mass
accretion rate changes by several orders of magnitude over a period
of week, or shorter (Gladstone, Done \& Gierli\'nski 2007). The X-ray
spectral evolution in the HID correlates with the X-ray variability
properties (Belloni et al. 2005) and with properties of the radio
emission from a jet (Fender et al. 2004).

\begin{table*}
\centering
\caption{Log of the {\it RXTE} observations. (1) The name of the
source. (2--4)
The year,
  and the date of the first/last observation we studied. (5) The
hydrogen column
density used in the spectral
  fits. (6--8) The distance (the uncertainty on the distance from the
literature), inclination and black hole
mass of the systems. (9) The number of analysed hard state
observations.
References for the entries in columns (5--8)
are as follows: [1] Done \& Gierli\'nski (2003), [2] Orosz \& Bailyn
(1997), [3]
Remillard \& McClintock (2006),
[4] Hynes et al. (2004), [5] Zdziarski et al. (2004), [6]
Munoz-Darias, Casares
\& Martinez-Pais (2008),
[7] since the inclination to \gx\ is unknown we assume
30$^{\circ}$, [8]
Kolehmainen \& Done (2010).}
\begin{tabular}{l c c c c c c c c}
\hline
Source & Year & Start & End & N$_{\rm H}$ & D & $i$ & M & Number \\
& & MJD$^a$ & MJD$^a$ & $\times$10$^{22}$ cm$^{-2}$ & kpc &  &
M$_{\odot}$ &
\\
(1) & (2) & (3) & (4) & (5) & (6) & (7) & (8) & (9)\\
\hline
\gro & 2005       & 53423 & 53673 & 0.8 [1] & 3.2 (3.0--3.4) [1] &
70$^{\circ}$ [2] & 6.3 (6.0--6.6) [3] & 94\\
\gx & 1996--1999 (o1)& 50291 & 51441 & \multirow{3}{*}{$\Biggr\}$ 0.6
[1]} &
\multirow{3}{*}{10 (6--15) [4--6]} & \multirow{3}{*}{30$^{\circ}$ [7]}
&
\multirow{3}{*}{10 (6.2--15) [6, 8]} & \multirow{3}{*}{224}\\
    & 2002/2003 (o2) & 52367 & 52774 & & & & & \\
    & 2004/2005 (o3) & 53044 & 53537 & & & & &\\
\hline
\end{tabular}\\
\begin{minipage}[b]{\linewidth}
$^a$ Modified Julian Date, ${\rm MJD} = {\rm JD} - 2400000.5$
\end{minipage}
\label{tab:tab1}
\end{table*}

\begin{figure}
\includegraphics[height=8.3cm,bb=188 69 400
758,clip,angle=-90]{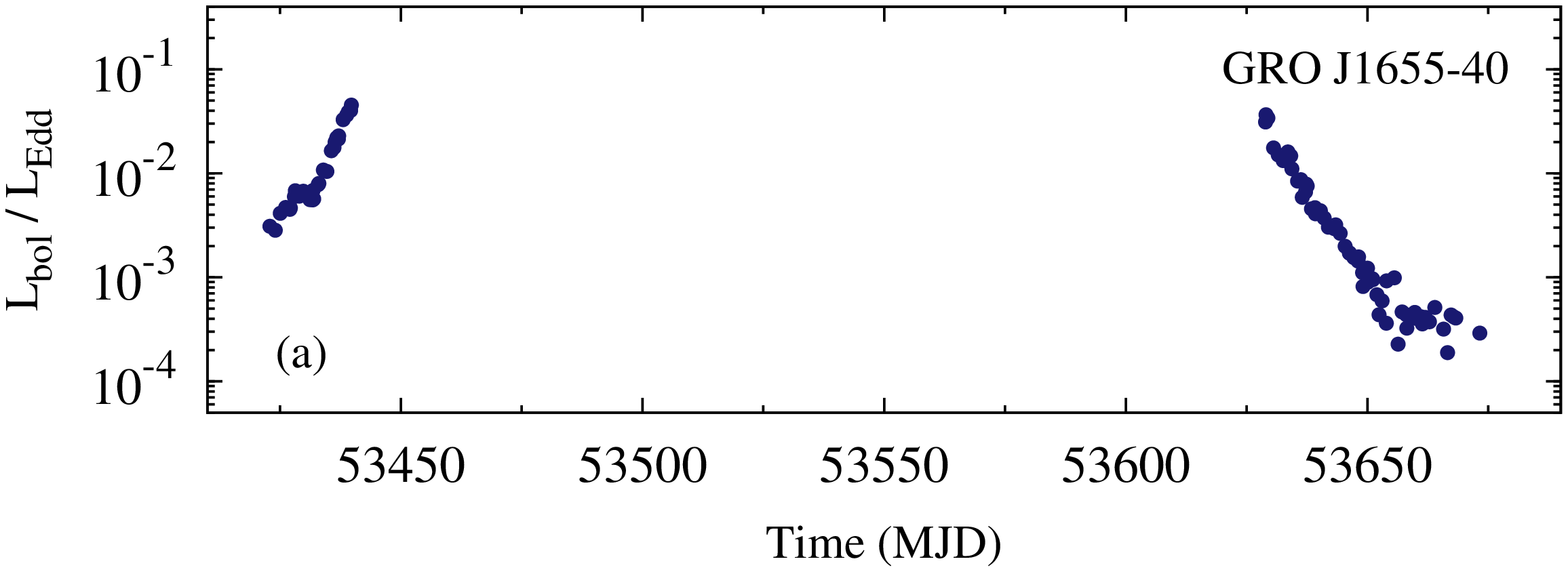} \\ 
\includegraphics[height=8.3cm,bb=188 69 440
758,clip,angle=-90]{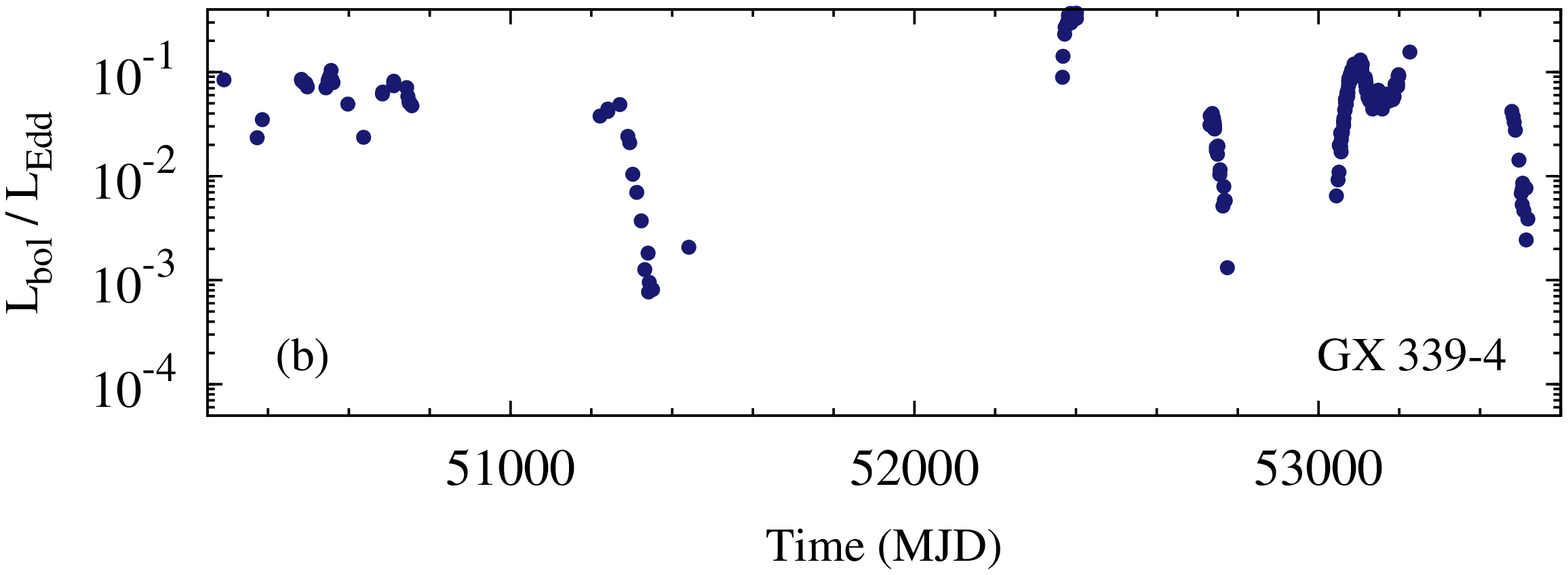} 
\caption{Hard state lightcurves of (a) \gro\ and (b) \gx. See e.g.
Done et al. (2007)
for complete lightcurves. Bolometric luminosities in Eddington units
have been
calculated based on the best fitting models of Sobolewska et al.
(2011, \gro)
and Model 2 in this work (\gx) extrapolated to the 0.01--1000 keV
band.}
\label{fig:fig1}
\end{figure}

\begin{figure*}
\centering
\includegraphics[height=5.8cm,angle=-90]{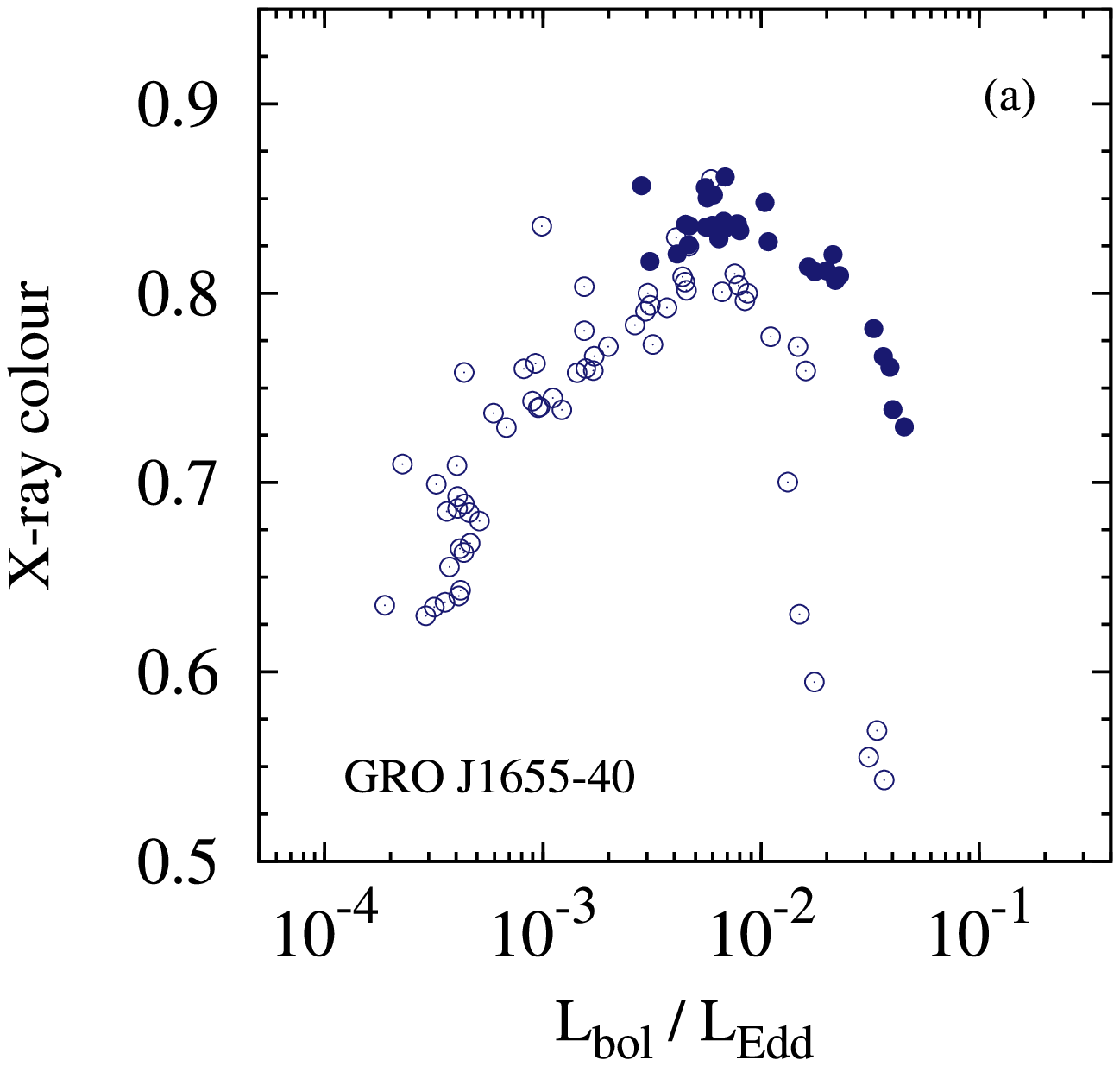} 
\includegraphics[height=5.8cm,angle=-90]{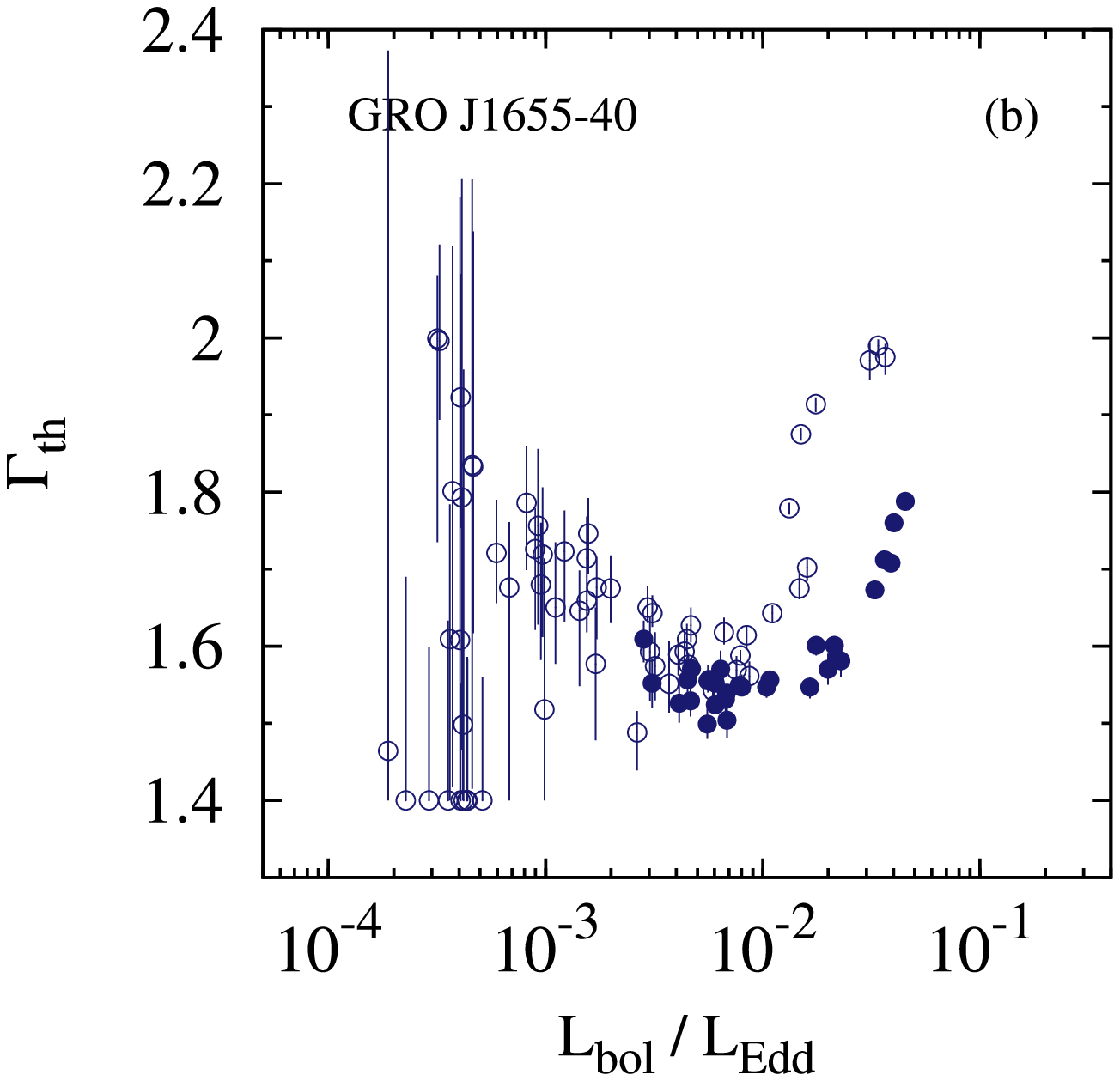} 
\includegraphics[height=5.8cm,angle=-90]{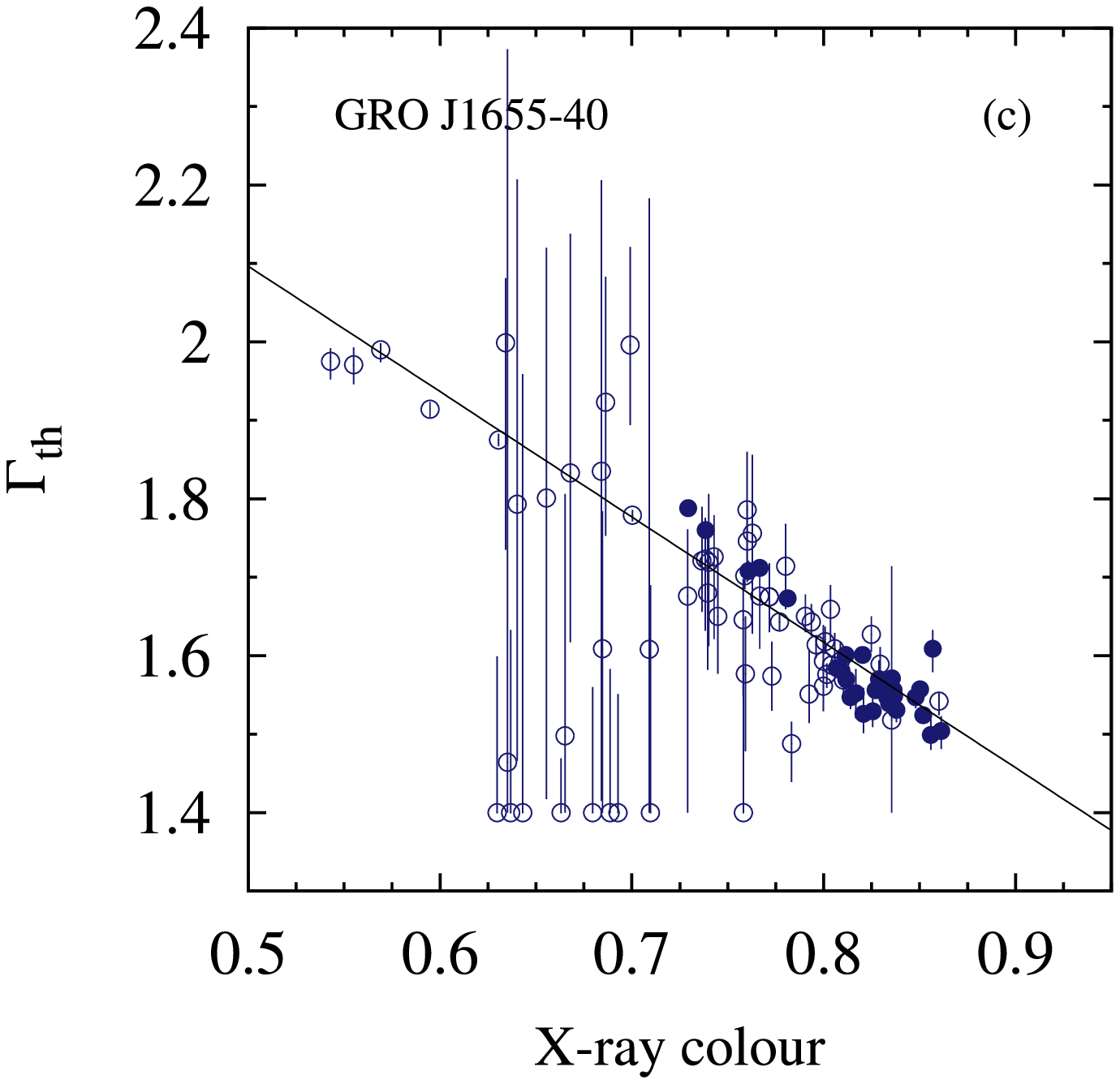}\\ 
\includegraphics[height=5.8cm,angle=-90]{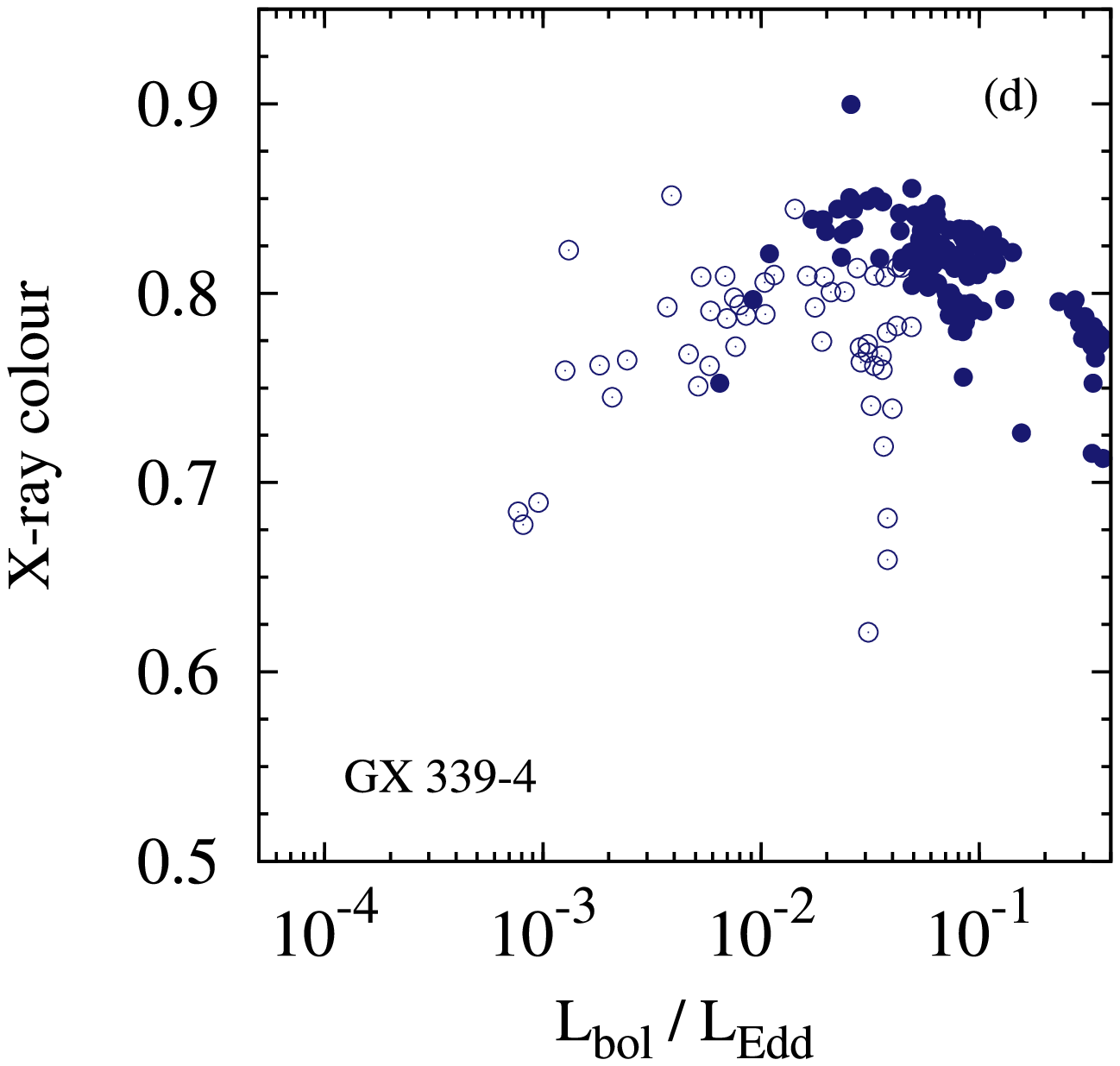} 
\includegraphics[height=5.8cm,angle=-90]{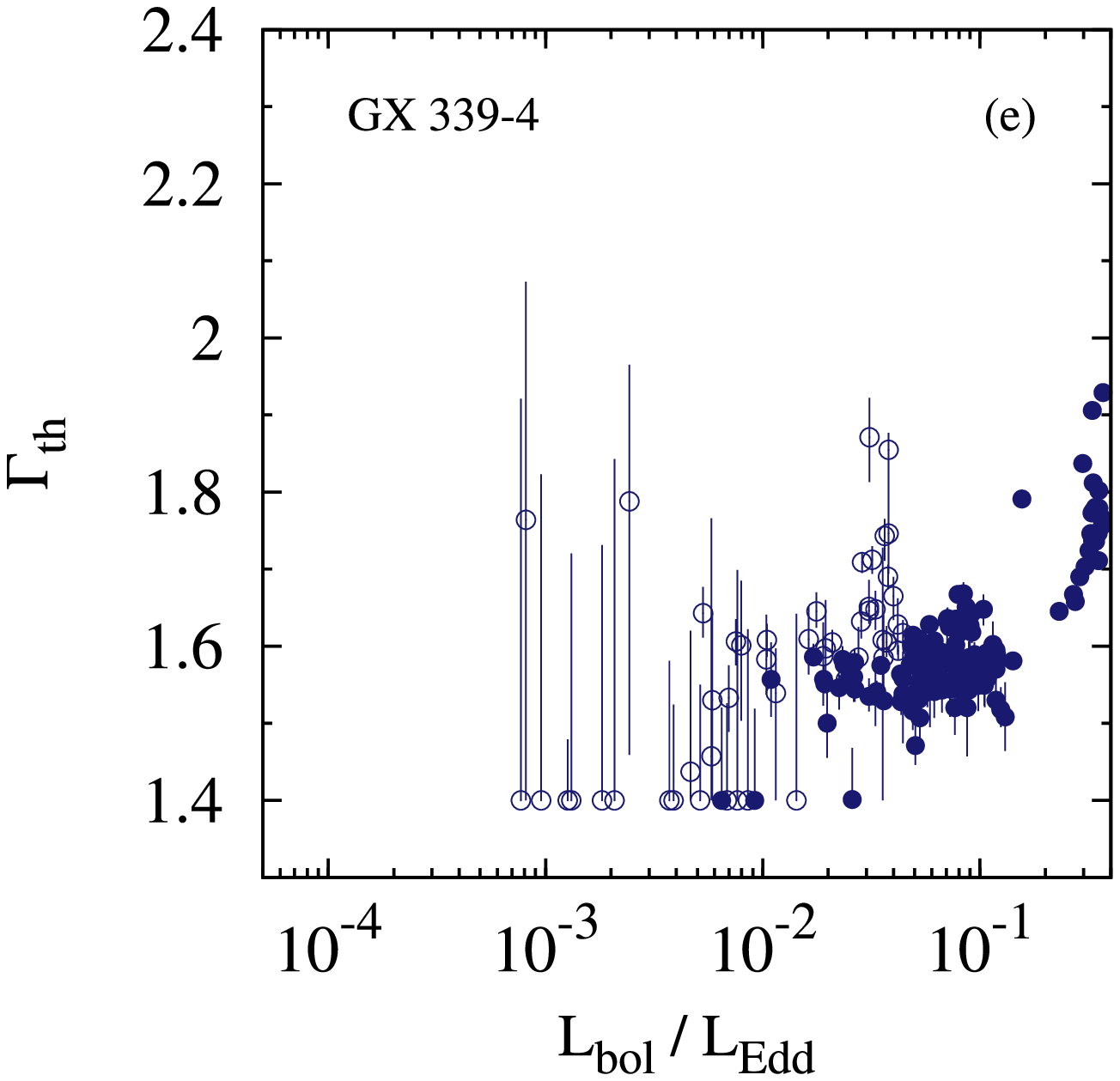}
\includegraphics[height=5.8cm,angle=-90]{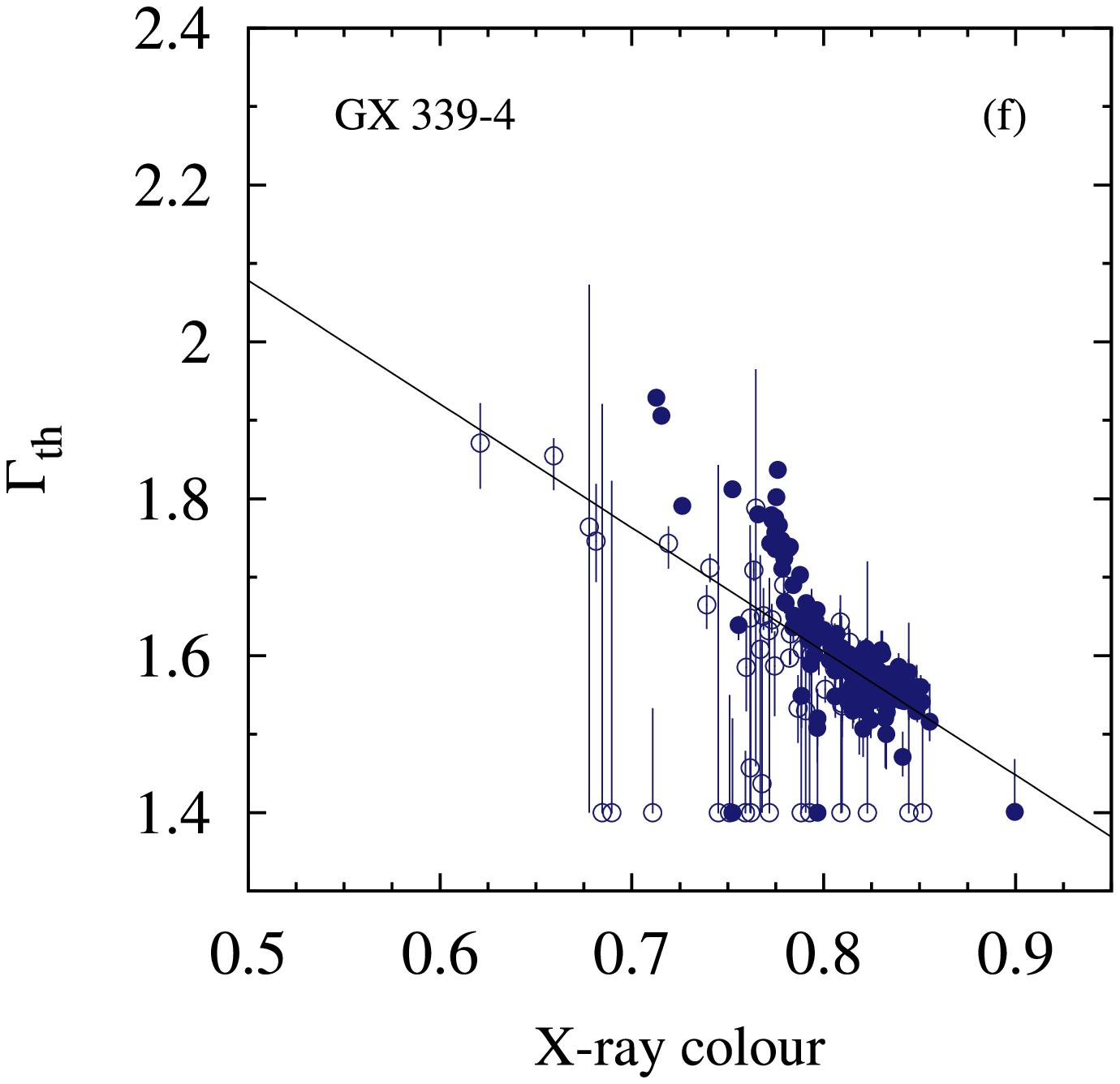}\\
\caption{(a/d) The hard state evolution of the X-ray colour with the
luminosity in Eddington units in  \gro\ and \gx; filled symbols --
rise,
opened symbols -- decay. A hysteresis pattern can be seen in
both sources. (b/e) The X-ray photon index as a function of the
luminosity in Eddington units (the best fitting photon indices are
those reported by Sobolewska et al. 2009). (c/f) The correlation
between the X-ray colour and the photon index suggesting that in \gro\
and \gx\ the colour variations are caused by the photon index
variations in the hard spectral state at all luminosity levels. The
solid line represents the best fitting linear function; in \gx\ we fit
only the decay data (see text). The observations that give only upper
limits on the photon index were neglected during the fit.}
\label{fig:fig2}
\end{figure*}

In the HID an increase in intensity is associated with an
increase in
the mass accretion rate, while a decrease in hardness corresponds to softening
of the X-ray photon index.
A number of scenarios were presented to explain the spectral evolution of GBHs,
i.e. the spectral softening (hardening) with increasing
(decreasing) mass accretion rate (e.g. reviews in Remillard \&
McClintock 2006; Done et al. 2007). One
of the scenarios is that of the truncated accretion disc (e.g. Esin et
al. 1997). In this model at low mass accretion rates the standard
accretion disc in the hard state is truncated far away from
a black hole and a hot inner flow is formed close to the black hole.
In such
geometry, few disc photons can be intercepted by the hot flow and the resulting
spectra have hard photon index, characteristic of the hard state. As the
mass accretion rate increases, the inner radius of the disc moves closer to the
black hole resulting in an increased cooling of the electrons. Consequently, the
spectrum softens and the source evolves to a soft state. The soft
state spectra seem to be well understood in terms of the multicolour
disc black body emission from an untruncated (extending to the innermost
stable circular orbit) disc accompanied by a weak Comptonised hard
tail.
Alternative hard state models not involving a truncated disc were
proposed (see Done et al. 2007 for strengths and weaknesses of each model).
These include e.g. a patchy outflowing corona above a disc extending to the last
stable orbit (Beloborodov 1999a; Malzac et al. 2001) and the jet origin of the
hard X-rays (Markoff, Falcke \& Fender 2001).

However, several recent studies have shown that in the hard state
at very
low mass accretion rates, below a few per cent of the Eddington rate, the
spectra of GBHs begin to soften again while the luminosity
decreases (e.g. Ebisawa et al. 1994; Revnivtsev, Trudolyubov \&
Borozdin 2000; Corbel et al. 2004; Jonker et al. 2004; Kalemci et al.
2005; Wu \& Gu 2008; Dunn et al. 2010). Interestingly, a similar
spectral evolution with luminosity has been  reported also in the case
of AGN. Above a certain luminosity, the photon index softens with
increasing luminosity (e.g. Porquet et al. 2004; Shemmer et al. 2006;
Saez et al. 2008; Sobolewska \& Papadakis 2009), but at low accretion
rates the opposite trend is observed: the photon index hardens while
the luminosity increases (Constantin et al. 2009; Gu \& Cao 2009),
just like in GBHs.

As mentioned, it is generally believed that the hard power-law like X-rays are
produced from Compton up-scatter of soft disc photons by energetic electrons in
a hot corona located close to the black hole.
The high energy cut-offs observed in the hard state spectra
suggest that the population of electrons is thermal (e.g. Gierli\'nski et al. 1999;
Rodriguez, Corbel \& Tomsick 2003; Miyakawa et al. 2008;
Joinet, Kalemci \& Senziani 2008; Motta, Belloni \& Homan 2009).
Within this framework, the main parameter that determines the spectral shape of the 
intrinsic hard X-ray continuum in accreting sources is the ratio of
heating-to-cooling
compactnesses, \lhls, where compactness is defined as a dimensionless luminosity, 
\begin{equation}
\ell = \frac{L}{R}\frac{\sigma_T}{m_ec^3},
\end{equation}
where $L$ is the luminosity of a spherical region of radius $R$, and $\sigma_T$ is the
Thomson cross section. 

In this paper we study how the compactness ratio, \lhls, 
evolves with GBH luminosity in the hard state. We consider the rise and decay parts of the
outbursts of two confirmed black hole binaries, when their luminosity is less
than $\sim 0.1$--0.3 of their Eddington limit. 
We fitted their spectra with a disc black-body
component (to account for their disc emission in soft X-rays) and the {\sc
eqpair} model of Coppi (1999)
to determine \lhls. Since this parameter is mainly determined by the
disc/corona geometry, our aim is to infer and constrain the evolution
of the source geometry in the hard state, and investigate which of the current
theoretical models are consistent with our results.

\section{Data selection}
\label{sec:data}

In this work, we re-analyse all the archival RXTE observations of
\gro\ and
\gx,  available up to 2007. Both sources are well studied systems
containing a dynamically confirmed black hole primary (see references in
Tab. 1). The \gro\ selected data cover in great detail the period of its 2005
outburst. In addition, the mass and distance to \gro\ are relatively well
constrained (Tab. 1; but see Foellmi 2009, and Sec. 4.3), which
allows for accurate conversion of the observed fluxes into the Eddington
luminosity ratios. The \gx\ selected data extend over a time-scale of several
years. They cover reasonably well its 2002/2003 and 2004/2005 outbursts (o2 and
o3, respectively) and include observations taken in the period of 1996--1999 (o1).
Even though the mass and
distance constraints of \gx\ are not as tight as those of \gro, we choose this
system for our study due to a large number of hard state observations displayed
in its multiple outbursts. Both sources follow a well known q-shaped pattern in
the colour-intensity diagram (e.g. Dunn et al. 2010), which means that their
X-ray spectral properties are typical of GBHs. Hence, we believe that the
data we selected are representative of a GBH behaviour during the luminosity
rise and decay phases of an outburst.

We extracted archival Standard 2 spectra for detector 2, top layer, in
the 3--20 keV
band and HEXTE spectra from both detectors in the 20--200 keV band. The details on data
reduction are given in Sobolewska, Gierli\'nski \& Siemiginowska (2009). Standard methods were used
to subtract the background, create response matrices and deal with systematic errors.
We obtained one spectrum per pointed observation. 

Of all the available RXTE observations, we chose to study the data
when the
source's high energy spectrum has photon index $\Gamma < 2$, as found from
the PCA+HEXTE 3--200 keV model fits including thermal Comptonisation component
(Sobolewska et al. 2009). Figure~\ref{fig:fig1} shows the X-ray
luminosity in
Eddington units plotted as a function of time, for the
selected RXTE observations of the two sources. The rising/decaying phases of the
outbursts can be clearly seen. In all cases, the
source luminosity was lower than 10\% of the Eddington limit for \gro\
and 30\% for \gx. Consequently, both systems during these
observations
should be mainly in their hard state. Luminosity estimates are based on the
Model 2 that we describe in the next section. In total, we studied 94 and
224 hard state observations of \gro\ and \gx, respectively.

In Tab. 1 we list the time period covered by the selected data, the hydrogen
column density
used in the spectral fits, the physical properties (i.e. distance, inclination and black
hole mass) we adapted in this work, and the number of the hard state
observations that we studied. The distance to \gro\ and its mass are known
quite accurately (but see Foellmi 2009). However, the estimates
for \gx\ suffer from significant uncertainties, so in this work we addopted
the canonical values (for GBHs) of 10 M$_{\odot}$ and 10 kpc for this source.

\section{Spectral modeling}
\label{sec:models}

The process of Comptonisation of the soft disc photons in the hot
corona
is widely accepted as the mechanism responsible for the bulk of the power-law like
continuum in the X-ray spectra of accreting objects. We
considered two different models to fit the 3--200 keV band spectra of our two sources.
We included a multicolour disc blackbody ({\sc diskbb} in {\sc Xspec}) in both of them to account
for the disc emission. As for the power-law continuum, we used two different 
Comptonisation models: the {\sc thcomp} thermal Comptonisation routine
of
Zdziarski, Johnson \& Magdziarz (1996), and the {\sc eqpair} code  of Coppi (1999).
In both models we fixed the temperature of the soft disc photons at $kT_{\rm bb} = 0.4$
keV because the RXTE low energy bandpass does not provide enough coverage
to measure properly the temperature of the cool hard state discs. We modeled the
spectra using {\sc Xspec} ver. 11.3.2 (Arnaud 1996).

\subsection{Estimating the photon index}

Sobolewska et al. (2009) modeled the 3--200 keV PCA/HEXTE
data of several GBHs. They used a model consisting of {\sc diskbb}, {\sc thcomp}, a  Gaussian line
profile to model iron K$\alpha$ features, and a smeared edge ({\sc smedge}) to account
for iron K$\alpha$ absorption. The complete model was described in {\sc Xspec}
as {\sc constant*wabs*smedge*(gaussian + diskbb +
thcomp)}, hereafter Model 1.  The N$_{\rm H}$ for the absorption
component {\sc wabs} was kept fixed to the Galactic column density listed in Tab. 1.
A constant component  was added to account for differences between the PCA and
HEXTE normalisations (the constant was fixed at 1 for PCA and was left
free to
vary for HEXTE). We used the Sobolewska et al. (2009) results to study the evolution
of the photon index in \gro\ and \gx.

\subsection{Estimating the heating-to-cooling compactness ratio}

Sobolewska, Siemiginowska \& Gierli\'nski (2011) modeled the hard state \gro\
observations of the 2005 outburst replacing the {\sc thcomp} model with {\sc eqpair},
a hybrid thermal/non-thermal Comptonisation model (Coppi 1999, Gierli\'nski et al. 1999).
The main parameter determining the spectral shape in the {\sc eqpair} model is the
ratio between the compactness of seed
photons, $\ell_s$, and hot electrons, $\ell_h$. This ratio, \lhls, defines the
spectral shape of the Comptonised continuum. It is a physical
parameter (as opposed to
the photon index in Model 1) and it depends mostly on
the geometry of the accretion flow. Typically, the hard state spectra are
characterised by $\ell_h/\ell_s \gg 1$, soft state spectra by
$\ell_h/\ell_s
\ll 1$ and very high/intermediate state spectra by $\ell_h/\ell_s \sim 1$.  
The complete model was defined in {\sc Xspec}
as {\sc constant*wabs(diskbb+eqpair)}, hereafter
Model 2, where {\sc constant} and {\sc wabs} were defined as in Model 1.

As described in Sobolewska et al. (2011), 8 of 94 hard state observations
showed residuals around 6--7 keV reminiscent of an iron line, and so the {\sc laor}
relativistic line model (Laor 1991) was included to improve these fits ({\sc
eqpair} does not account for the iron line emission). With this addition, Model 2 fits
to all 94 hard state datasets of \gro\ resulted in a null hypothesis probability higher than
0.05.

In this
paper we applied the same model to the selected hard state observations of \gx. 
We found that 48 of 224 datasets required an addition of the iron line, and
again we added the {\sc laor} component to Model 2. Following this, the model 
provided a good fit (with a null hypothesis probability higher than 0.05)  to
220 of the 224 datesets. We therefore concluded that  Model 2 is a
good description of the hard state spectra of this source as well.

We use the results of Sobolewska et al. (2011) for \gro\ and our results for
\gx\ to study the hard state evolution of \lhls.

\subsection{Estimating the bolometric luminosity}

The spectra we studied are characterised by a photon
index lower than $\sim$2. We assume that they originate in a process of
thermal Comptonisation, and so they peak in the $EF_E$ representation
at energies of the order of the temperature of the thermal electron. Indeed, many
of our spectra show a roll over at energies of the order of 100 keV.
In order to estimate the total X-ray flux, and ensure
that we do not neglect any energy radiated above 100 keV, we extrapolated the best fitting
Model 2 spectra, and estimated the flux in the 0.01--1000 keV band, $F_{\rm 0.01-1000}$. Since
the majority of energy in GBHs is radiated in the X-ray band, we considered 
$F_{\rm 0.01-1000}$ to be representative of the bolometric flux. Assuming isotropic
emission, we defined bolometric luminosity as $L_{\rm bol} = 4\pi D^2 F_{\rm 0.01-1000}$,
for a black hole located at a distance $D$. We study the spectral evolution of our sources
as a function of bolometric luminosity in Eddington units, $L_{\rm bol}/L_{\rm E}$,
where $L_{\rm E} = 1.3\times 10^{38}$
M/M$_{\odot}$ ergs s$^{-1}$ for a black hole of mass M.

\section{The spectral evolution in the hard state}
\label{sec:results}

\begin{figure*}
\centering
\includegraphics[width=4.9cm,bb=176 169 528
541,clip,angle=-90]{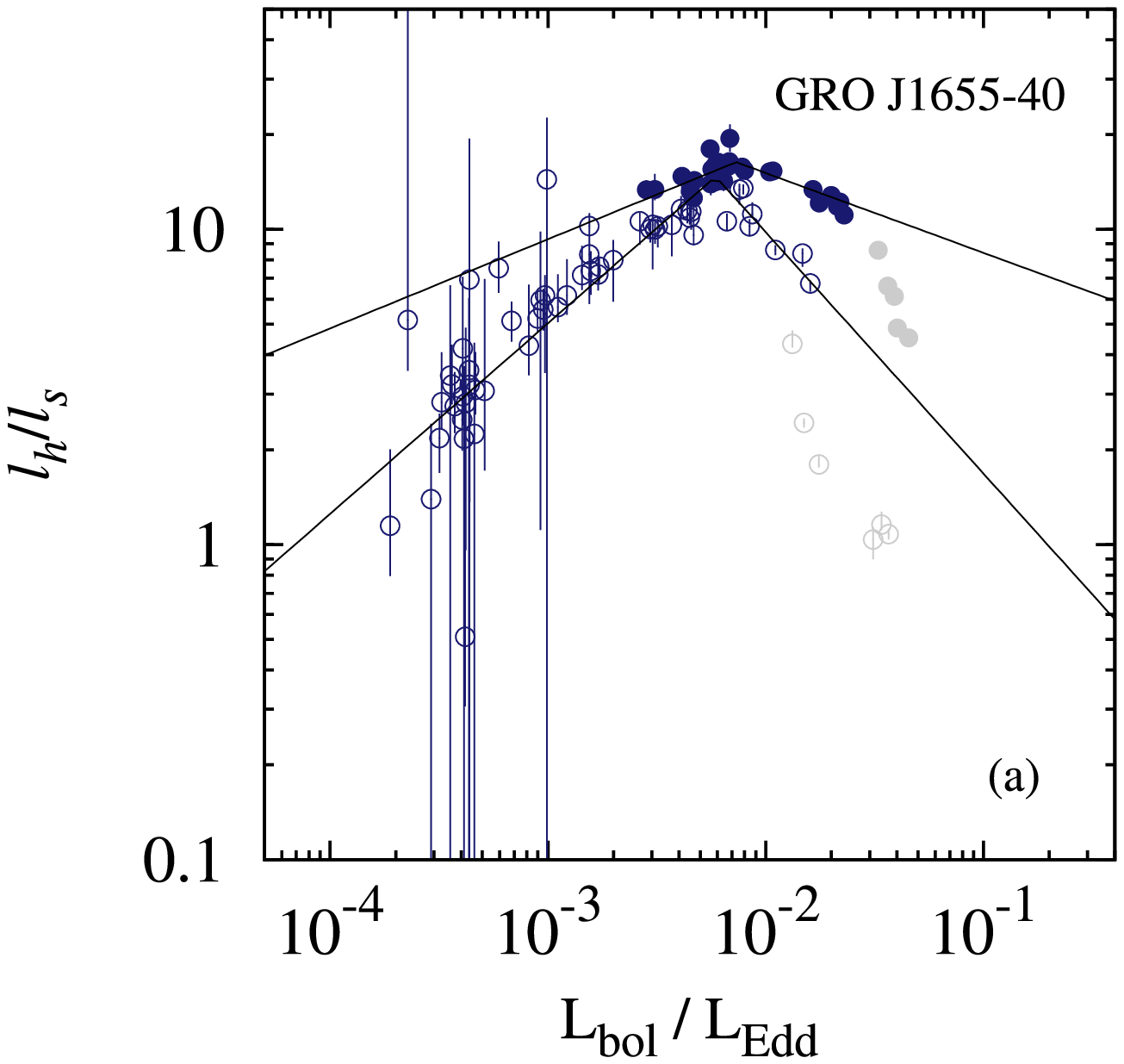}
\includegraphics[width=4.9cm,bb=176 255 528
541,clip,angle=-90]{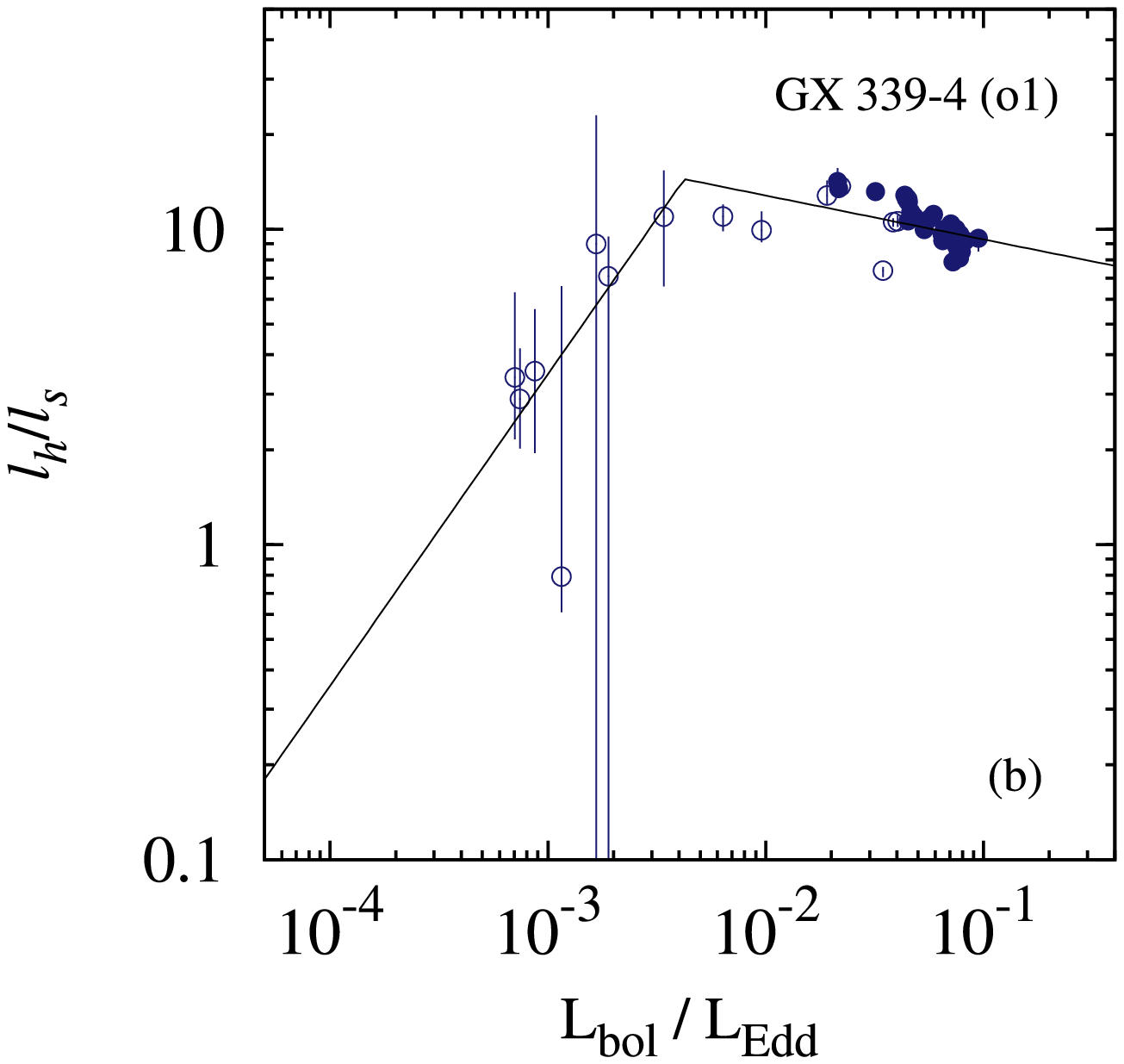}
\includegraphics[width=4.9cm,bb=176 255 528
541,clip,angle=-90]{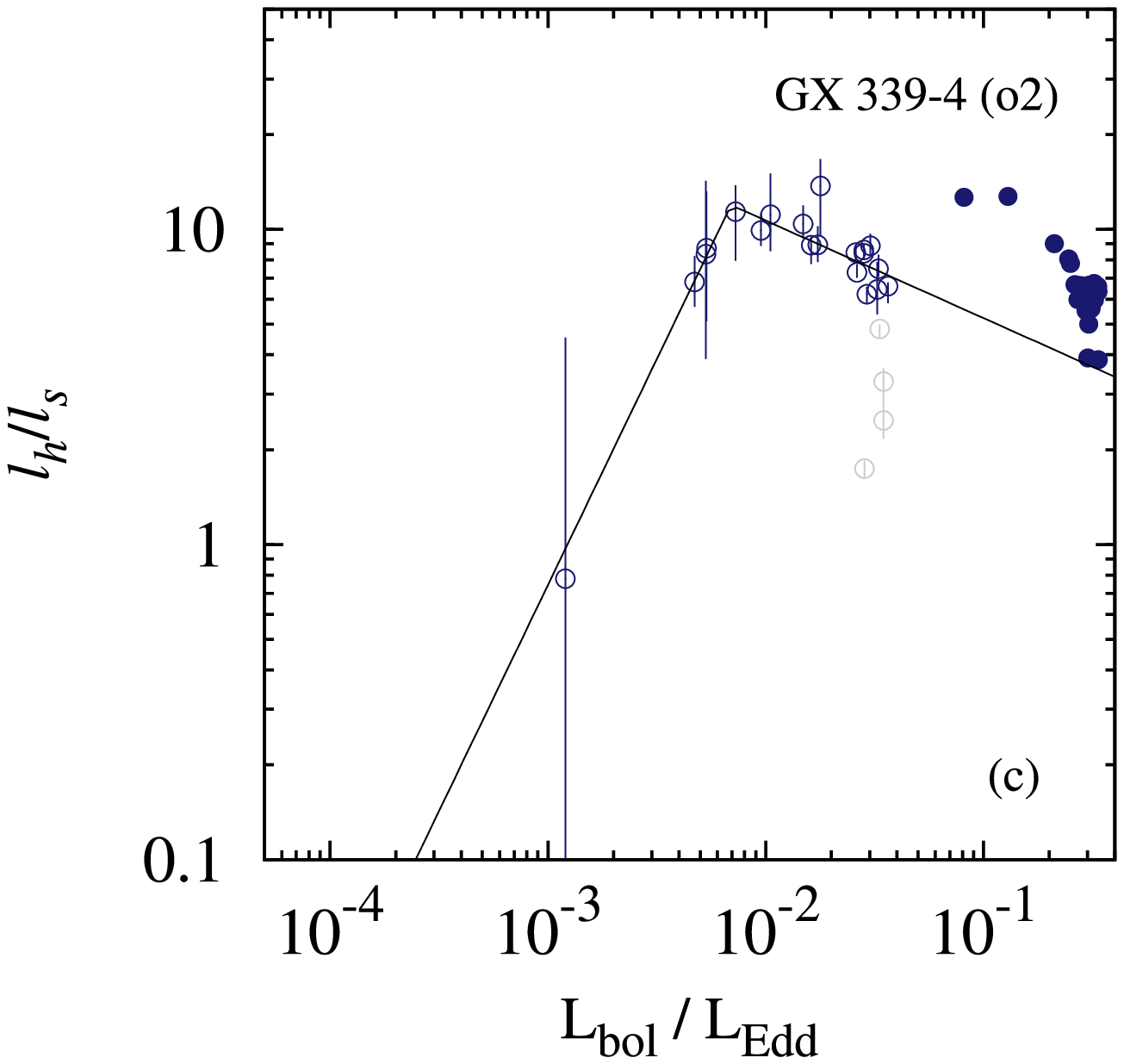}
\includegraphics[width=4.9cm,bb=176 255 528
541,clip,angle=-90]{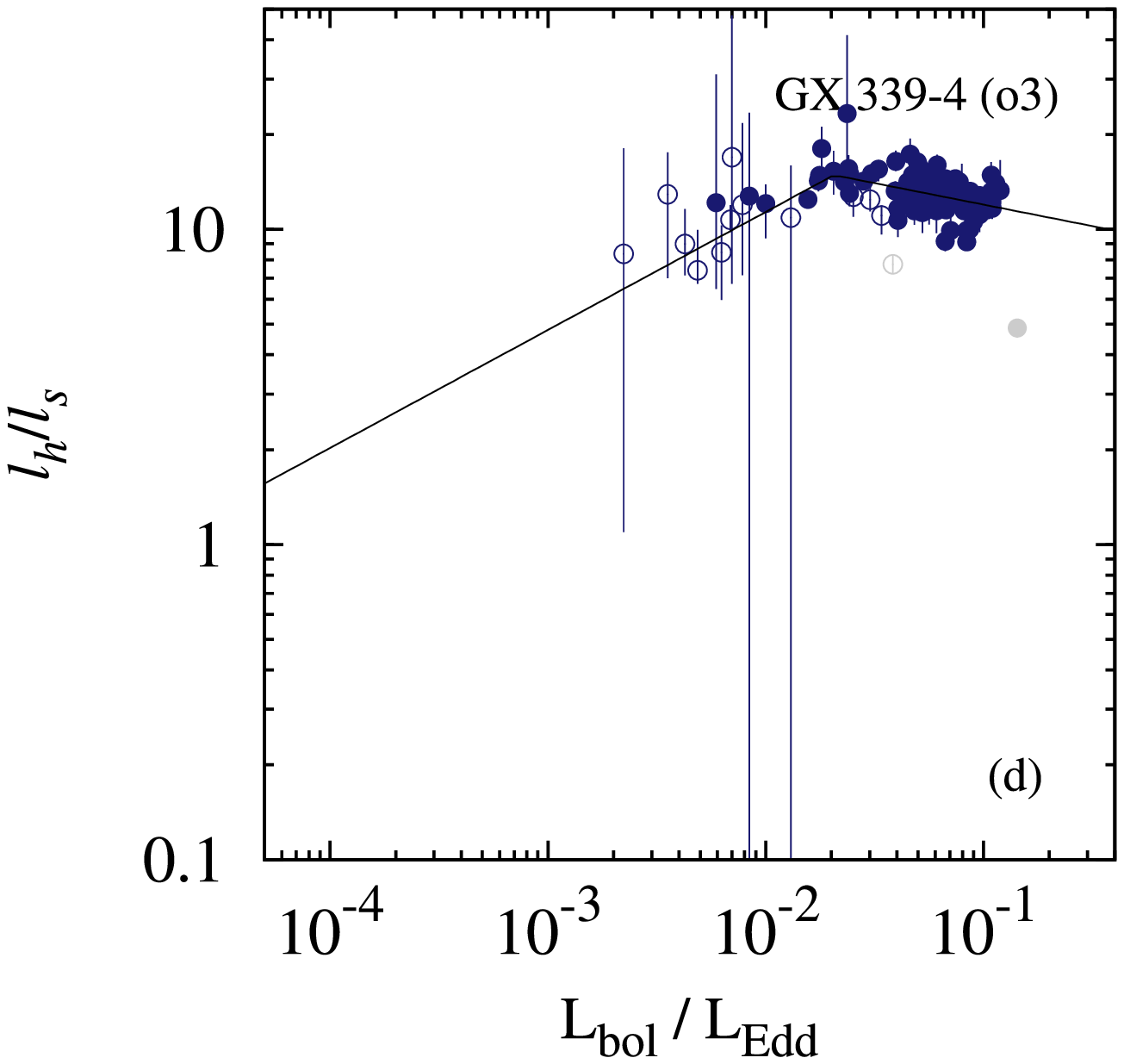}
\caption{The hard state heating-to-cooling compactness ratio of the
corona, \lhls\, as a function of the luminosity in Eddington units for
(a) \gro\ and (b--d) \gx. Filled symbols -- rise; open symbols --
decay. Solid lines indicate the broken power law
fits to the data. The light gray data points were neglected during the
fit (see Sec. 4.3).
In \gx\ the best fitting break luminosity is $L_{\rm crit,GRO} =
(0.0061\pm0.0002)L_{\rm E}$,
and in \gx\ the average of the fits to the three outbursts is $L_{\rm
crit,GX} =
(0.011\pm0.002)L_{\rm E}$.}
\label{fig:fig3}
\end{figure*}

\subsection{X-ray colour evolution}
\label{sec:color}

We first considered the evolution of the X-ray colour with source accretion
rate in order to study the spectral variability of the two sources in a model independent
way, without any a priori assumption about the shape of the X-ray continuum.
Following  Belloni et al. (2005), we defined  the X-ray colour as the ratio of the
observed count rates in the hard, i.e.  6.3--10.5 keV, and soft, i.e. 3.8--6.3 keV,
X-ray energy bands. Our results are shown in
Figs.~\ref{fig:fig2}a/d. Note that, traditionally, this kind of plots show a
count rate as a function of X-ray colour. However, we chose to plot the
Eddington luminosity ratio for a more straightforward comparison of the plots in
Figs.\ref{fig:fig2}a/d with the following plots in the present work.

The figures show clearly that the X-ray spectra of \gro\ and \gx\ vary with
luminosity during the hard state. The X-ray colour follows a well defined
pattern, which is similar in both sources.  It first increases, i.e. the hard
X-rays become more dominant, with increasing luminosity until  $L_{\rm bol}/L_{\rm E}$
reaches the value of $\sim 1$--3\%, above which  the colour decreases. This
value obviously depends on the adopted distance and black hole mass estimates for each
source, however the inferred  spectral evolution with luminosity cannot be
attributed to any specific model assumptions because the colour was calculated
using the observed count rates. We indicate the rising and decaying
phases of the outbursts with different symbols and notice the
hysteresis pattern in both sources.

\subsection{Photon index evolution} 
\label{sec:thcomp}

Although the X-ray colour is a model-independent measure of the spectral shape
of the sources, it is not easy to interpret its evolution with source
luminosity, as colour variations can be caused either by a variable 
normalisation
of the disc and Comptonisation components, and/or by intrinsic
variations of the
hard band photon index. For that reason,
Figs.~\ref{fig:fig2}b/e shows a plot
of the best fitting X-ray photon index from Model 1 as a function of source luminosity,
for the same observations that are plotted in Figs.~\ref{fig:fig2}a/d. The
$\Gamma$ values plotted in these panels correspond to the Model 1 best fitting
results of Sobolewska et al. (2009).

The best fitting $\Gamma$ values for the lowest flux spectra have rather large
errors associated with them. In fact, in some cases (specially in \gx), the best
fit value reached the lowest boundary allowed during the fit, set to $\Gamma=1.4$
(Sobolewska et al. 2009). In these cases, the photon index
estimates should be considered as upper limits. These effects
complicate the determination of the photon index evolution with
luminosity, at the lowest flux states of the
sources.

Overall, we believe that, the evolution of  $\Gamma$ with
$L_{\rm bol}/L_{\rm E}$ as shown in  Figs.~\ref{fig:fig2}b/e, shows a pattern
which is similar to that of X-ray colour evolution (Figs.~\ref{fig:fig2}a/d). As
the source luminosity increases, $\Gamma$ becomes harder, until $L_{\rm bol}/L_{\rm
E}$ reaches the level of a few per cent, above which the spectrum softens with
increasing luminosity. This result suggests that, to a large extend, the X-ray
colour evolution is caused by intrinsic photon index variations when GBHs are in their
hard state. To strengthen this conclusion we plot the photon
index as the function of the X-ray colour in Figs.~\ref{fig:fig2}c/f. 
It is apparent that the two quantities correlate and their dependence
can be modeled with a linear function.
During the fit we neglect the data points being upper limits (but we include them in
Figs.~\ref{fig:fig2}c/f), and we use the average of the upper and lower
$\Gamma$ errorbars (from Model 1) as the error on $\Gamma$.
We obtain $\Gamma_{\rm GRO} = (-1.60\pm0.04)\times {\rm C}+(2.90\pm0.03)$,
$\Gamma_{\rm GX} = (-1.6\pm0.2)\times {\rm C}+(2.9\pm0.2)$, where C stands for the X-ray
colour. The fits result in relatively high $\chi^2/d.o.f.$ of 3.8 and 2.3 for \gro\ and \gx,
respectively. Nevertheless, it is clear that the model describes rather well the observed
anti-correlation. In the case of \gx\ we fit only the data points of the
outburst decay. The observations taken during the rising phase with ${\rm C}>0.8$
are consistent with the best fit. Those with the X-ray colour lower
than $\sim$0.8 (and $L > 0.2 L_{\rm E}$) seem to deviate from the best fitting line.
However, the photon index and colour anti-correlation holds also for them, with perhaps
a different slope due to relatively high luminosity of these hard
state spectra.

\subsection{Correlation between \lhls\ and luminosity}

\begin{figure}
\centering
\includegraphics[height=7.0cm,bb=176 167
469 550,clip,angle=-90]{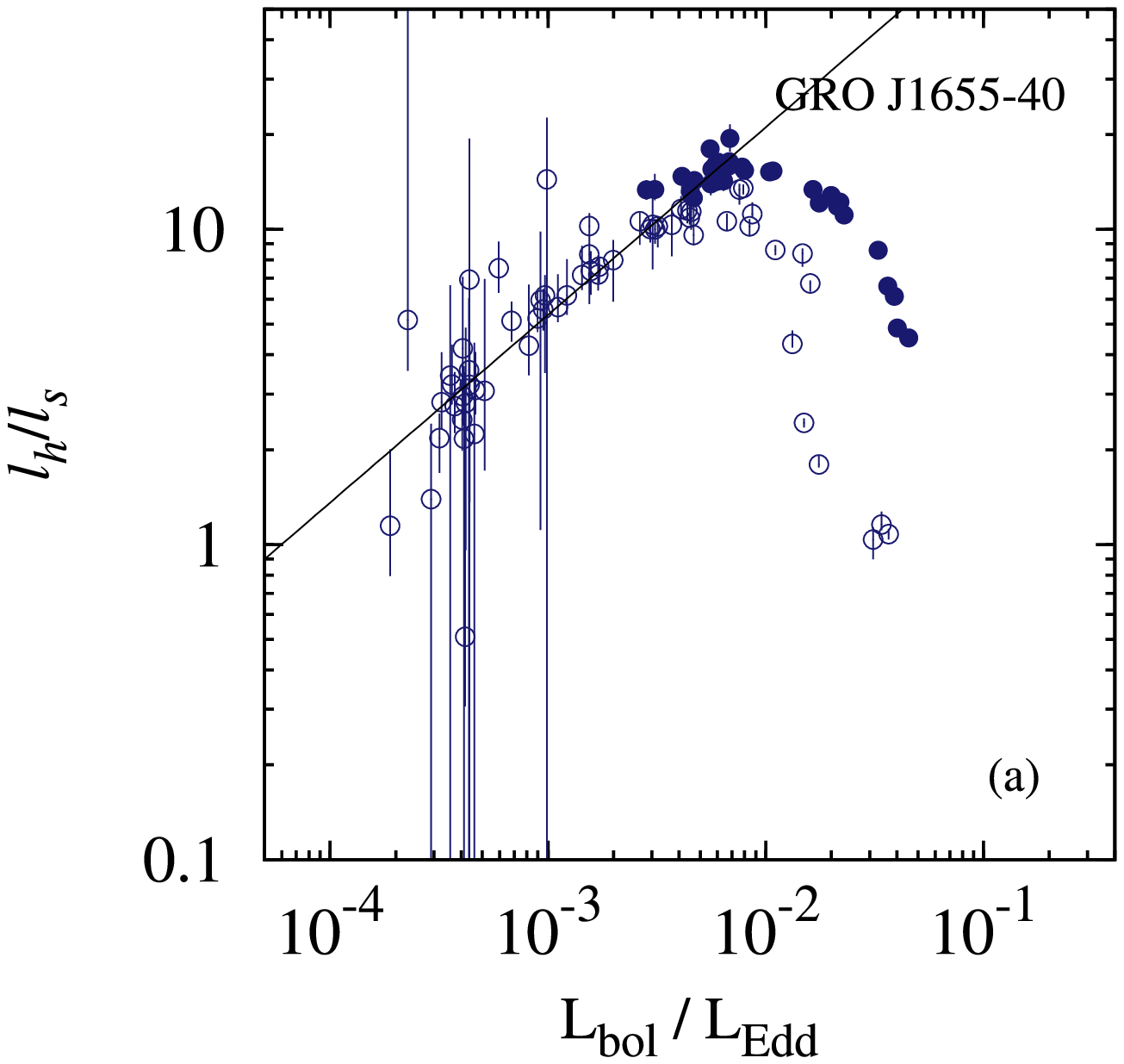}\\
\includegraphics[height=7.0cm,bb=176 167 532
550,clip,angle=-90]{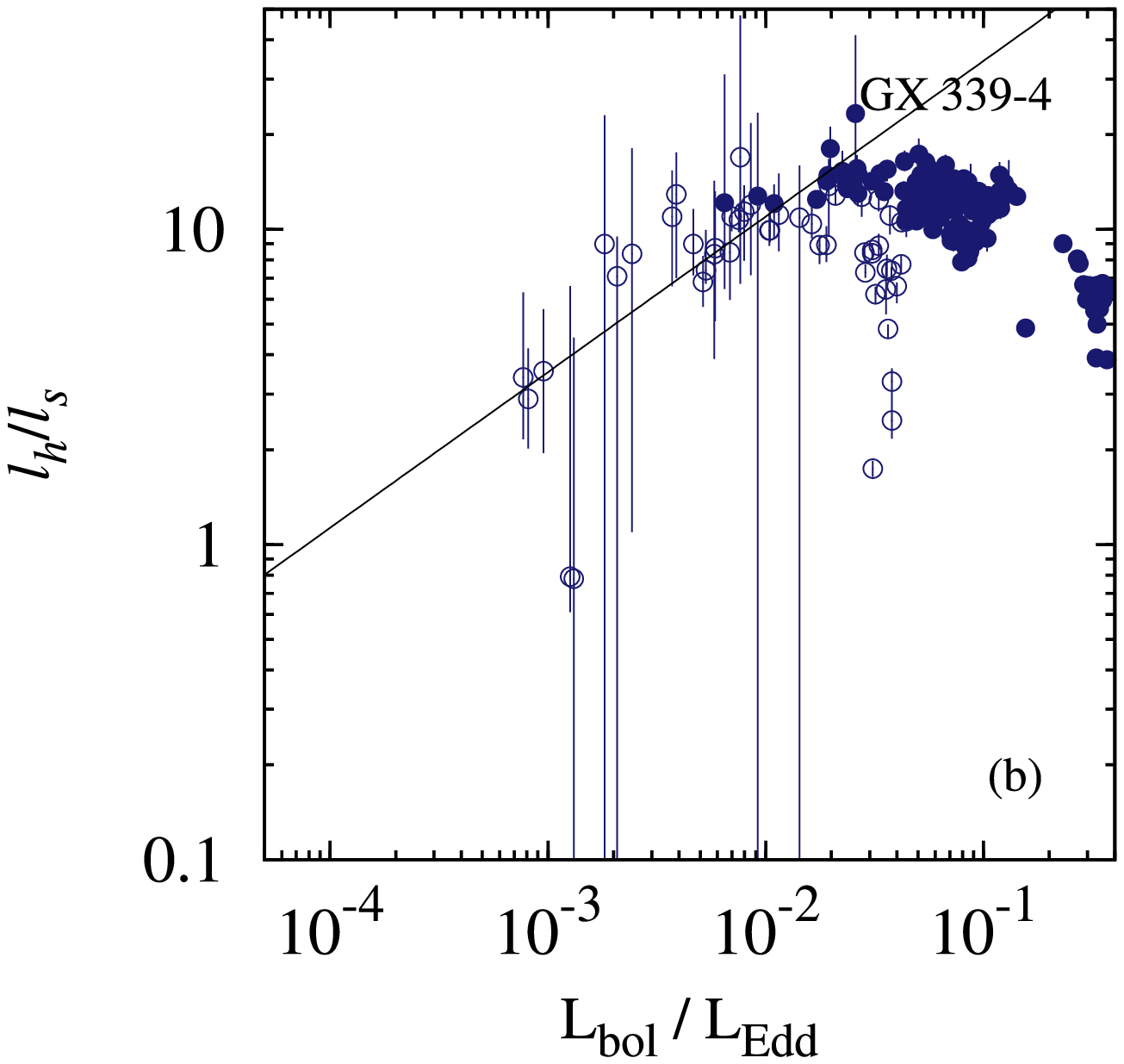}
\caption{The hard state heating-to-cooling ratio of the corona, \lhls\, as a function of the
luminosity in Eddington units for (a) \gro\ and (b) \gx. Filled
symbols -- rise; open symbols -- decay. Solid lines indicate the
best fitting model to the data with $L_{\rm bol}$ below $L_{crit}$
(see Sec. 4.3). Best fitting
slopes are  $a_{\rm GRO}=0.60\pm0.03$ and $a_{\rm GX}=0.49\pm0.15$.}
\label{fig:fig4}
\end{figure}

Having fitted the hard state spectra of the two sources with Model 2, we can
now investigate the observed spectral variations in a physically meaningful
way, by studying variations of the hard-to-soft compactness ratio with luminosity.
For typical coronal temperatures, the ratio of the hard-to-soft compactnesses
determines the photon index of the hard X-ray
continuum (e.g. Beloborodov 1999b). Consequently, under the hypothesis of inverse-Compton produced
X-rays, the observed spectral shape variations should be caused by the  \lhls\
variations, which are associated mainly with the changes of the accretion geometry,
e.g. the inner edge of the disc proceeding towards the black hole, or the
presence of a coronal outflow with variable velocity.

We present the evolution of our best fitting \lhls\ values  as a function of
luminosity in Fig.~\ref{fig:fig4}. These plots clearly indicate that
the observed  decrease of the photon index, and the increase of
the X-ray colour with increasing luminosity up to a few percent of the
Eddington
limit, is due to the fact that \lhls\ increases from a value of $\sim 1$ to
$\sim 10$--15 for the same range of $L_{\rm bol}/L_{\rm E}$, in both sources. At
Eddington luminosity ratios higher than a few per cent the \lhls\ evolution changes.
It decreases with luminosity leading to softer spectra (and lower values of X-ray colour).

In order to find the critical value of luminosity at which the
compactness
ratio reaches maximum, we fit a broken power law model to the \lhls\ vs. $L_{\rm bol}/L_{\rm E}$
data of the two sources and summarize the fit results in Tab. 2.
Similarly to the case of the photon index and X-ray colour correlation,
we neglect measurements that are upper limits on \lhls\ (but we include them in the plots),
and use the average of the lower and upper \lhls\ errorbars (from Model 2) as the error on \lhls.
In the case of \gro\ we fit separately the data from
the rise and decay and we calculate their weighted average, which gives
$L_{\rm crit,GRO} = (0.0061\pm0.0002)L_{\rm E}$.
In the case of \gx\ we treat each outburst separately, and we fit the rise and decay
together for outbursts o1 and o3. We use only the decay data of outburst o2 because
the data of the rise do not allow to put meaningful constraints on the critical luminosity,
and a strong hysteresis effect prevents a joint fitting of the rise and decay data.
Again, we calculate the weighted
average of the three measurements, and we get $L_{\rm crit,GX} = (0.011\pm0.002)L_{\rm E}$.
During the fits we neglect several
data points indicated in Fig.~\ref{fig:fig3} with light gray symbols. These observations correspond to
the periods shortly before/after the hard-to-soft/soft-to-hard spectral transition, when
the compactness ratio, as well as the X-ray colour and photon index, change
rapidly with luminosity.

However, the most significant contribution to the uncertainty on $L_{\rm crit}$ is introduced by the
uncertainty of the GBH distance and mass estimates.
Taking into account these uncertainties in the case of \gx\ (Tab. 1) we derive
$0.003 L_{\rm E} < L_{\rm crit, GX} < 0.04 L_{\rm E}$,
while in the case of \gro\ we get much tighter constrains,
$0.005 L_{\rm E}< L_{\rm crit, GRO} < 0.007 L_{\rm E}$.
Foellmi (2009) derives an
upper limit to the \gro\ that differs substantially from the accepted
value of 3.2 kpc. Adopting their upper limit of 2 kpc and ${\rm M}=6.3{\rm M}_{\odot}$
we obtain $L_{\rm crit, GRO} = 0.002 L_{\rm E}$.
However, Caballero Garc{\'{\i}}a et al. (2007)
pointed out that if \gro\ were indeed located so close, its companion star would
not fill its Roche lobe.
Given all the uncertainties, and the range of possible values for $L_{\rm crit}$ that we
presented above, it seems quite possible that $L_{\rm crit}/L_{\rm E}$ is the same
in both objects, and roughly equal to $\sim$0.01.

Having determined the luminosity at which the evolution of \lhls\ changes, we
proceeded to determining the slope of the \lhls\ vs. $L_{\rm bol}/L_{\rm E}$ positive
correlation below $L_{\rm crit}$.  We fitted jointly the rise and
decay data with luminosities below $L_{\rm crit,GRO}$ and $L_{\rm
crit,GX}$, respectively,
with the following function:
$\ell_h/\ell_s= b \times (L_{\rm bol}/L_{\rm E})^a$. The model describes well
the anti-correlation between \lhls\ and luminosity ($\chi^2/d.o.f.$
equals 1.5 and 0.7 in \gro\ and \gx, respectively). The solid lines
in Fig.~\ref{fig:fig4}a/b indicate the best fitting models.
The best fitting values are:  $a_{\rm
GRO}=0.60\pm0.03$, $a_{\rm GX}=0.49\pm0.07$, $b_{\rm GRO}=328\pm 47$, $b_{\rm
GX}=107\pm39$ (Tab. 2). They are consistent within their errors
for the two sources, and their weighted means are $\bar{a}=0.58\pm 0.02$
and $\bar{b}=197\pm 30$.

\begin{table}
\centering
\caption{Coefficients in the \lhls\ vs. $L_{\rm bol}/L_{\rm E}$ correlation.}
\begin{tabular}{l c c c}
\hline
Source & $a$ & $b$ & $L_{\rm crit}/L_{\rm E}$\\
\hline
\gro  & $0.60\pm0.03$ & $328\pm47$ & $0.006$\\
\gx   & $0.49\pm0.07$ & $107\pm39$ & $0.01$\\
\hline
\end{tabular}\\
\label{tab:tab2}
\end{table}

\subsection{Quiescence}

\begin{figure}
\centering
\includegraphics[height=7.0cm,bb=176 167 532 550,clip,angle=-90]{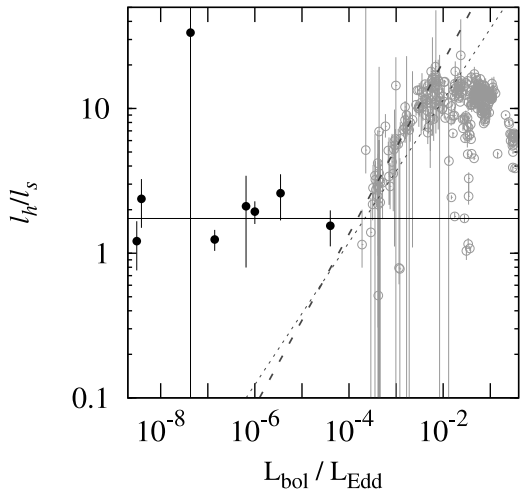}
\caption{Comparison of the hard state and quiescent state. Data and
fits
to \gro\ and \gx\ are as in Fig. 4 (open/gray symbols, dashed and
dotted lines, respectively). The filled/black symbols represent the measurements
for several GBHs collected from the literature (see Sec. 5.5). In quiescence the
\lhls\ saturates at $\sim$1.7 (solid horizontal line).}
\label{fig:fig5}
\end{figure}

The anti-correlation between \lhls\ and Eddington luminosity ratio does not
extend to the quiescence. This conclusion is based on
the Eddington luminosity ratios and photon indices collected for quiescent
accreting GBHs by Corbel, Tomsick \& Kaaret (2006, XTE~J1118+480, A0620-00, XTE~J1550-564,
\gx, \gro, V641 Sgr). We included also
three measurements for V404 Cyg. In two cases, we combined the reported photon indices with the
Eddington luminosity ratios calculated based on the 0.3--10 keV
XMM-Newton unabsorbed flux (Bradley et al. 2007) and 3--9 keV Chandra flux
(Corbel, Koerding \& Kaaret 2008), assuming ${\rm M} = 12{\rm M}_{\odot}$ and $D=3.5$ kpc
(Corbel et al. 2006). The third
quiescent observation of V404 Cyg was originally published by Kong et al.
(2002). However, it was claimed to be affected by pile-up, so we addopted the
pile-up corrected photon index of Corbel et al. (2008), and the Eddington
luminosity ratio as in Kong et al. (2002).

These quiescent measurements of
the photon index are consistent with being constant,
$\Gamma_{\rm q} = 2.13\pm0.06$, as a function of luminosity.
In Fig. 5 we plot again the hard state data of
\gro\ and \gx\ together with \lhls\ calculated based on the quiescent
photon indices using the following equation, $\Gamma = 2.33 \times (\ell_h/\ell_s)^{-1/6}$
(Beloborodov 1999b). We conclude that in quiescence the heating-to-cooling compactness
ratio saturates at $\ell_h/\ell_s \sim 1.7$, corresponding to
$\Gamma_{\rm q}$.

\section{Discussion}

We studied two confirmed GBHs, namely \gx\ and
\gro, considering all their archival \rxte\ observations until June 2007.
We focused on a subset of observations, when the photon index was
smaller
than 2, i.e. when the sources were mainly in the hard state. During these
observations, the source luminosity was lower than 10\% of the Eddington limit in
the case of \gro, and 30\% in \gx.

We used traditional X-ray colours and the results of Sobolewska et al. (2009) to determine the
spectral evolution of each source. Our results are consistent with what has been
observed in the past, for other GBHs and AGN as well: the X-ray colour
increases, and the photon index {\it hardens},  with increasing
luminosity up to the point where $L_{\rm bol}/L_{\rm E}\sim 0.01$. Above this point,
the opposite trend is observed: a decrease of the X-ray colour and a softening
of the photon index with increasing luminosity.

Within the context of thermal Comptonisation models for the X-ray
production in
accreting black holes, the photon index is determined by the
hard-to-soft
compactness ratio, \lhls. The main result of our work is that the observed
spectral variations of the two sources, when in the hard state, can be
explained if \lhls\ and luminosity are anti-correlated for luminosities
below $L_{\rm crit}\sim 0.01L_{\rm E}$, and positively correlated for luminosities
higher than $L_{\rm crit}$. The anti-correlation below $L_{\rm crit}$ can be
described by $\ell_h/\ell_s \propto L_{\rm bol}^{~\bar{a}}$,
where $\bar{a} = 0.58\pm 0.02$ is the weighted mean for both objects.

Since the compactness ratio, \lhls, depends mainly on the geometry of the accretion flow,
our results can put constrains on various models that have been proposed to
explain the hard X-ray emission in GBHs. We discuss below some of these constrains
and implications of our results. We note that in the hard state spectra of GBHs
the Comptonised emission dominates over the emission of the seed photons, so that
$L_{\rm bol} \propto \ell_h$.

\subsection{Hot thermal inner flow, truncated disc seeds}

The hard X-rays could be produced by Compton up-scattering of seed
photons from a disc illuminating a hot inner flow.
Heat conduction from thermal contact of the hot flow and cool disc can
lead to evaporation of the inner thin disc, resulting in a truncated thin disc,
with a radius of $R_{\rm disc}\propto \dot{m}^{-1/2}$ (Czerny,
R\'o\.za\'nska \& Kuraszkiewicz 2004).
We assume that the hot flow is powered by the same mass accretion rate
that flows through the thin disc, so that its luminosity is given by
the remaining potential energy from the inner
edge of the truncated disc, $R_{\rm disc}$, to the inner edge of the hot
flow, $R_{\rm in}$.  Then (approximately) $$\ell_h\sim
GM\dot{m}(1/R_{\rm in}-1/R_{\rm disc})\times
\eta_{\rm corona}/\eta_{\rm disc},$$ where
$\eta_{\rm corona}/\eta_{\rm disc}$ is the radiative
efficiency of the hot flow relative to that of a thin disc, and is
proportional to $\dot{m}$ for a radiatively inefficient flow
(e.g. Sharma et al. 2007), and roughly constant for a radiatively efficient flow.
If we assume that $R_{\rm disc}\gg R_{\rm in}$, then $\ell_h \propto
\dot{m}\eta_{\rm corona}/\eta_{\rm disc}$, which implies that
$\ell_h \propto\dot{m}^2$ for a radiatively inefficient flow, and $\ell_h \propto\dot{m}$
for a radiatively efficient flow. The truncated disc luminosity is
$L_{\rm disc}=GM\dot{m}_{\rm disc}/2R_{\rm disc}$, but only part of this is
intercepted by the hot Comptonisation region, which we assume to be a sphere
of radius $R_{\rm h}$. This fraction, $f$, varies with radius,
and reaches maximum at the truncation radius of the disc,
where $f=(R_{\rm h}/R_{\rm disc})^2/\pi$. Since the
disc luminosity peaks at the same radius, the seed photon luminosity
for the hot flow is
$$\ell_s\approx 
GM\dot{m}/2R_{\rm disc} \times (R_{\rm h}/R_{\rm disc})^2/\pi,$$
so that
$$\ell_h/\ell_s  = 2\pi \times (R_{\rm disc}/R_{\rm in}) \times (R_{\rm disc}/R_{\rm h})^2 \times \eta_{\rm corona}/\eta_{\rm disc},$$
and we get that
$ \ell_h/\ell_s\propto \dot{m}^{-3/2}\eta_{\rm corona}/\eta_{\rm disc}
\propto \ell_h^{~-1/4}$ for a radiatively inefficient flow, and proportional
to $\ell_h^{~-3/2}$ for a radiatively efficient flow.  Either way, this
predicts that $\ell_h/\ell_s$ increases as the luminosity decreases,
i.e. that the spectrum hardens at lower luminosities. This is 
due mainly to the rapidly decreasing solid angle subtended by the hot
flow to the truncated disc. However, this is in
marked contrast to the observed {\em softening} of the spectrum seen
at the lowest luminosities, followed by the saturation in
quiescence, ruling out this scenario as the physical mechanism for
the observed X-ray emission at luminosities below $L_{\rm crit}$.

However, this could instead explain the observed 
behaviour above $L_{\rm crit}$, where the spectrum abruptly softens
with increasing luminosity. Hence, there must be a transition to another
mechanism at $L_{\rm crit}$ to explain the change in \lhls\ behaviour.

\subsection{Hot thermal inner flow, cyclo-synchrotron seeds}

There are also seed photons generated in the hot flow itself, by
cyclo-synchroton radiation of the energetic electrons. This emission is
strongly self absorbed for thermal electrons, so it peaks at
the self absorption frequency $\nu_{\rm ssa}$, given where the
cyclo-synchrotron emissivity equals that of a blackbody.  The full
equation for this is complex (Narayan \& Yi 1995). However,
Wardzi\'nski \& Zdziarski (2000) give a numerical approximation for
this frequency, $\nu_{\rm ssa}\propto\theta^{0.95}\tau^{0.05}B^{0.91} $, where
$\theta$ and $\tau$ are the dimensionless temperature and optical depth of the
electrons in the hot flow, and $B$ is the magnetic field.  At the
observed temperatures, $\nu_{\rm ssa}$ should be on the Rayleigh-Jeans
tail of the blackbody, so the flux at this point is $\propto \nu_{\rm ssa}^2
\theta$. Thus the seed photon luminosity $\ell_s\propto \nu_{\rm ssa}^3
\theta$. The hard luminosity
is as before, with $\ell_h=GM\dot{m}/2R_{\rm in}\times
\eta_{\rm corona}/\eta_{\rm disc}$, hence $\ell_h/\ell_s \propto
\ell_h/(\theta^{3.85}\tau^{0.15}B^{2.73})$.

We get a further constraint as $\ell_h/\ell_s$ is also set by $\theta$
and $\tau$. Pietrini \& Krolik (1995) give an approximation for this
of $0.1(\ell_h/\ell_s)^{1/4}= \theta\tau$.
This gives $\ell_h/\ell_s\propto \ell_h^{~0.51}\tau^{1.19}
(\tau/B^2)^{0.70}$. In a radiatively inefficient flow, the
gas pressure is set by the ion temperature, which remains
approximately constant around the virial temperature. Hence
$\tau/B^2$ is constant if the magnetic
pressure is a constant fraction of the gas pressure.
Hence $\ell_h/\ell_s\propto \ell_h^{~0.51}\tau^{1.19}\propto
\ell_h^{~1.1}$ if $\tau\propto\dot{m}\propto \ell_h^{~1/2}$.
This is somewhat faster than the observed  $\ell_h/\ell_s\propto
\ell_h^{~0.5-0.6}$, but given the level of approximation this
at least describes the observed trend below $L_{\rm crit}$
for the spectrum to soften as the luminosity decreases.

However, this alone does not give a framework for the change in
behaviour at higher luminosities, where the spectrum dramatically
softens with increasing luminosity. Thus neither seed photons from the
disc nor self produced synchrotron self Compton photons from the flow
can explain the full range of observed behaviour, though the
combination of the two (seed photons from the disc changing  to
seed photons from cyclo-synchrotron as luminosity decreases) can
explain the observations.

The saturation observed in quiescence is not expected if the
electron distribution in the hot flow is thermal. However, it is likely
that the electron distribution is not a pure Maxwellian. Evidence for
the presence of a non-thermal high energy tail in the electron distribution
was found in the hard state of several black hole binaries (McConnell et al. 2002;
Wardzi\'nski et al. 2002; Cadolle Bel et al. 2006; Joinet et al. 2007;
Droulans et al. 2010). Consequently, in quiescence, while the plasma becomes optically
thin (due to decreasing $L/L_{\rm E}$), the Comptonised component
becomes dominated
by the contribution arising through scattering on the non-thermal high-energy electrons,
and the evolution of $\Gamma$ (and hence \lhls) depends on the microphysics of particle
acceleration in the corona. Particle acceleration with a slope that
does not depend on $L/L_{\rm E}$ would result in a saturation of
\lhls.

\subsection{Outflowing hot thermal corona, untruncated disc}

An outflowing hot corona above an untruncated disc is an alternative
geometry to explain the observed hard spectra seen in the hard
state (Beloborodov 1999a; Malzac et al. 2001).
Here the seed photons for Compton scattering are from the disc, but are suppressed
by relativistic beaming from the mildly outflowing corona.

In the original model of Beloborodov (1999a), the intrinsic disc emission is assumed
to be negligible compared to reprocessed radiation produced by the
disc illumination by the X-ray source.  Then $\ell_h/\ell_s$ depends
mainly on the outflow velocity (as well as a few geometric
parameters).  As the velocity is increases, beaming reduces the
illumination of the disc and the soft photon flux returning to the
outflow. Therefore, as the velocity increases, the ratio $\ell_h/\ell_s$
increases, and the reflected fraction decreases. We also note that in the outflowing
corona model, the luminosity of the corona is amplified by relativistic
beaming and this affects the shape of the observed correlations
between $\ell_h/\ell_{s}$ and luminosity.

The observed spectral softening (and correlated increase in reflected
fraction: e.g. Gilfanov, Churazov \& Revnivtsev 1999; Ibragimov et al. 2005)
seen above $L_{\rm crit}$ can be explained in this model if the outflow
velocity (and hence $\ell_h/\ell_s$) decreases as luminosity increases.
The change in behaviour below $L_{\rm crit}$ then requires that
the outflow velocity decreases as luminosity decreases. This makes a clear prediction
that at the lowest luminosities where the beaming is negligible, the continuum should be
accompanied by a substantial reflected emission from an untruncated
disc and $\ell_{h}/\ell_{s}$ should saturate, as observed.

Alternatively, the softening of the spectrum below $L_{\rm crit}$ could
also be caused by an increase in intrinsic emission from the disc such
that this dominates over the reprocessed emission as the source of seed
photons. The outflow velocity could then remain high, or even continue
to increase as luminosity decreases. This predicts that the reflected
fraction remains small, but that the observed fraction of power dissipated
in the disc should increase.

\subsection{Non-thermal jet, untruncated disc}

Another type of emission suggested to explain the hard X-ray spectrum
is direct non-thermal synchrotron radiation from a jet (Markoff, Nowak
\& Wilms 2005). Both, thermal Comptonisation and jet models can
reproduce
the shape and observed energy range of the spectral cut-offs seen in the hard states
(e.g. Zdziarski et al. 2003; Markoff \& Nowak 2004).
In several black hole binaries the cut-off energy was reported to anti-correlate
with luminosity in a bright hard state (e.g. Wardzi\'nski et al. 2002;
Rodriguez, Corbel \& Tomsick 2003; Miyakawa et al. 2008; Joinet, Kalemci \& Senziani 2008;
Motta, Belloni \& Homan 2009).
This can be explained rather naturally in terms of thermal
Comptonisation models
by enhanced cooling by the disc photons. Thus, it seems likely that thermal
Comptonisation does dominate at $L_{\rm bol}/L_{\rm E}>0.01$.
It is then possible that the transition in spectral behaviour seen at
$L_{\rm crit}$ marks the point at which non-thermal emission from the jet
starts to dominate over thermal Comptonisation from the hot flow (Russell
et al. 2010), and theoretical models of jet dominated accretion flows
(e.g. Falcke \& Markoff 2000; Markoff et al. 2001; Yuan, Markoff \& Falcke 2002)
certainly predict that the kinetic power of the jet exceeds the radiative
luminosity (see also Malzac, Merloni \& Fabian 2004; Malzac, Belmont \& Fabian 2009).
However, the photon index of the X-ray emission then depends on
the photon index of the non-thermal electrons within the jet, so to
explain the spectral softening with decreasing luminosity requires that
the electron heating/cooling processes result in a steeper electron
distribution. There are currently no real constraints on this, due to
the lack of knowledge of jet physics.

In quiescence, similarly to the case of non-thermal Comptonisation
dominating
the X-ray in the scenario involving thermal inner flow and cyclo-synchrotron seeds,
the saturation of \lhls\ can be explained in the jet model if the slope
of accelerated non-thermal particles in the jet is independent on
$L/L_{\rm E}$.

\section{Conclusions}
We show that in the hard state of
GBHs the compactness ratio \lhls\ 
reaches maximum at $L_{\rm crit} \sim 0.01 L_{\rm E}$. For $L_{\rm
bol}<L_{\rm
crit}$,
we find that $\ell_h/\ell_s=\bar{b} \times (L_{\rm bol}/L_{\rm E})^{~\bar a}$,
where $\bar{a}=0.58\pm 0.02$ and $\bar{b}=197\pm 30$ are
the weighted slope and intercept based on the fits to the data
of \gro\ and \gx. We suggest that the change of the behaviour in the
\lhls\ vs. luminosity relation is caused by a change in the
X-ray emission
mechanism for the hard state spectra of GBHs. The relation does
not extend to
the quiescent state. Instead, a saturation of \lhls\ is observed at a
value of $\sim 1.7$.

The observed hard state evolution of \lhls\ can be explained if the seed photons
are cyclo-synchrotron
photons up to $L_{\rm crit}$, above which they are replaced with the truncated disc photons.
Alternatively, the observed variability of \lhls\ can be explained in the
case of an outflowing corona
if the outflow velocity increases with luminosity up to $L_{\rm crit}$ and then decreases
with increasing luminosity, or if the seed intrinsic disc photons are replaced with the reprocessed photons
above $L_{\rm crit}$.
Finally, in the scenario involving an X-ray emitting jet, the observed
transition
in \lhls\ behaviour may mark the point at which the non-thermal jet emission starts dominating over the thermal
Comptonisation emission.

Observed saturation of \lhls\ in quiescence can be explained in
terms of an untruncated disc
model with cyclo-synchrotron seed (and in the jet model) if the non-thermal particles in the
corona (jet) are accelerated with a slope that is independent on
$L/L_{\rm E}$. The outflowing
corona model naturally predicts the quiescent saturation of
\lhls\ if the outflow velocity
decreases with luminosity below $L_{\rm crit}$.

\section*{Acknowledgments}
We would like to thank the anonymous referee for helpful comments
and suggestions. MS acknowledges support by the Chandra grant GO8-9125A and
Marie-Curie ToK Fellowship number MTKD-CT-2006-039965. IEP acknowledges support
by the EU FP7-REGPOT 206469 grant. This work was partially supported by the GdR PCHE
in France.


\bsp

\label{lastpage}

\end{document}